\documentclass[preprintnumbers,superscriptaddress,showkeys]{revtex4}
\usepackage{amsmath,amsfonts,amssymb,amscd,amsxtra,amsthm}
\usepackage{graphicx}
\usepackage{xcolor}
\usepackage{bm}
\usepackage{orcidlink}
\usepackage{stackengine}
\usepackage{pgfplots}
\usepackage{booktabs}
\usepackage{array}
\usepackage{enumitem}
\usepgflibrary{plotmarks}
\usepackage{tikz-feynman}
\usepackage{subcaption}
\usepackage{float}
\usetikzlibrary{positioning}
\usepackage[justification=justified,singlelinecheck=false]{caption}
\begin{document}
\preprint{PKNU-NuHaTh-2026}
\title{Three-body decay $\phi\to\pi^+\pi^-\pi^0$ with Omnès-type final-state interactions}
\author{Seung-il Nam\,\orcidlink{0000-0001-9603-9775}}
\email{sinam@pknu.ac.kr}
\affiliation{Department of Physics, Pukyong National University (PKNU), Busan 48513, Republic of Korea}
\author{Jung Keun Ahn\,\orcidlink{0000-0002-5795-2243}}
\affiliation{Department of Physics, Korea University, Seoul 02841, Republic of Korea}
\date{\today}
\begin{abstract}
We investigate the decay $\phi\to\pi^+\pi^-\pi^0$ in an effective-Lagrangian framework that keeps the dominant $\rho\pi$ mechanism, the direct three-pion term, and a real constant background contribution explicitly separated at the amplitude level. The resonant contribution is described with the Gounaris-Sakurai propagator, the neutral channel includes $\rho^0$-$\omega$ mixing, and the leading elastic $I=1$, $P$-wave $\pi\pi$ final-state interaction is incorporated through the complex, $s$-dependent Omnès function $\Omega_1(s+i0)$. The Omnès factor is applied channel by channel to the three two-pion subsystems, while the direct amplitude is dressed by the averaged factor $\Omega_{\rm dir}=(\Omega_0+\Omega_++\Omega_-)/3$. To avoid double-counting the elastic rescattering already contained in the physical $\rho\to\pi\pi$ coupling, we introduce $g_{\rho\pi\pi}^{\rm bare}=g_{\rho\pi\pi}^{\rm phys}/|\Omega_1(m_\rho^2)|$ in the Omnès-dressed $\rho$-exchange amplitude. With the final parameter set used in this work, we obtain $\Gamma_{\rm th}=0.6915~{\rm MeV}$, slightly above the empirical estimate $\Gamma_{\rm exp}\approx0.660\pm0.020~{\rm MeV}$, while the direct integrated weight is reproduced at the KLOE scale, $I_{\rm dir}=8.393\times10^{-3}$. The corresponding resonant weight is $I_{\rho\pi}=0.9619$. The resulting Dalitz distribution retains the expected $\rho\pi$-dominated three-band structure and shows the localized neutral-channel deformation induced by $\rho^0$-$\omega$ mixing. Remaining discrepancies in the $x$ and especially the $y$ projections indicate the limitations of the present minimal single-channel Omnès implementation and motivate a next-stage analysis based on a full dispersive crossed-channel treatment and a direct fit to the efficiency-corrected KLOE Dalitz-bin data.
\end{abstract}
\maketitle
\section{Introduction}
The decay $\phi\to\pi^+\pi^-\pi^0$ is a particularly clean light-meson process for studying the coexistence of a dominant resonant mechanism and a subleading direct three-body contribution. Phenomenologically, the channel is governed primarily by the sequential process $\phi\to\rho\pi\to\pi^+\pi^-\pi^0$, which produces the characteristic three-band pattern associated with the $\rho^0\pi^0$, $\rho^+\pi^-$, and $\rho^-\pi^+$ subchannels in the Dalitz plot~\cite{KLOE:2003kas,ParticleDataGroup:2024cfk}. At the same time, anomalous effective interactions allow a direct coupling of the $\phi$ meson to the three-pion final state, and the possible size of this nonresonant component has long been discussed in the anomalous sector of low-energy QCD and in vector-meson-dominance descriptions of light-meson decays~\cite{Wess:1971yu,Witten:1983tw,Rudaz:1984bz,Kaymakcalan:1983qq,Fujiwara:1984mp}.

Experimentally, the Dalitz-plot analysis at the $\phi$ peak is sensitive not only to the resonance parameters of the charged and neutral $\rho$ mesons, but also to the magnitude and phase of a possible direct $3\pi$ amplitude in addition to the dominant $\rho\pi$ contribution. In the KLOE measurement of $e^+e^-\to\phi\to\pi^+\pi^-\pi^0$, the Dalitz-density distribution was used to separate the resonant $\rho\pi$ component, the direct term, and the much smaller $\omega\pi^0$ channel, yielding the integrated fractions $I_{\rho\pi}=0.937$, $I_{\mathrm{dir}}=8.5\times10^{-3}$, and $I_{\omega\pi}=2.0\times10^{-4}$, with the resonant-direct interference contributing at the level of about $6\%$. This pattern makes the decay a useful testing ground for examining how a small nonresonant amplitude can appear on top of a much larger resonant background~\cite{KLOE:2003kas}.

We also note several complementary approaches to the same physics. The decay $\phi\to3\pi$ has recently been analyzed with Khuri-Treiman dispersion relations, including crossed-channel effects and KLOE Dalitz-plot information~\cite{Garcia-Lorenzo:2025uzc}. Closely related three-pion dynamics have also been studied from lattice QCD through the extraction of the $\omega$-meson resonance parameters in a finite-volume three-body formalism~\cite{Yan:2024gwp,Mai:2017bge}. In addition, the decay width of $\phi\to3\pi$ has been evaluated in an NJL quark-model description of $\omega$-$\phi$ mixing~\cite{Volkov:2020jor}. These works provide useful complementary benchmarks for the present effective-Lagrangian treatment of the Dalitz distribution and final-state interactions.

A realistic phenomenological description requires at least three ingredients. First, because the physical Dalitz region spans a broad range of two-pion invariant masses, the intermediate $\rho$ meson should be described with a running-width representation; in the present work, we employ the Gounaris-Sakurai (GS) parametrization~\cite{Gounaris:1968mw}. Second, the neutral channel can receive a localized correction from $\rho^0$-$\omega$ mixing, familiar from VMD analyses of the timelike pion form factor and from isospin-violating effects in the vector-meson sector~\cite{McNamee:1974vb,Coon:1987kt,Bernicha:1994re,OConnell:1995nse,Gardner:1997ie}. Third, the dominant $\pi\pi$ rescattering in the isovector $P$ wave should be included at least approximately if one wants to discuss the overall normalization of the decay rate in a physically meaningful way.

Modern dispersive descriptions of $\omega/\phi\to3\pi$ incorporate analyticity and final-state interactions systematically and are indispensable for precision amplitude analyses~\cite{Niecknig:2012sj,Danilkin:2014cra}. The present work does not attempt a full dispersive reconstruction of $\phi\to3\pi$. Its aim is narrower: to construct a transparent effective-Lagrangian representation in which the resonant and direct pieces remain explicitly separated, while incorporating the leading elastic $\pi\pi$ rescattering effect in the $\rho$-dominated channel through an Omnès-type factor.

This point deserves emphasis. In the physical Dalitz region, the decay is governed primarily by the sequential mechanism $\phi\to\rho\pi\to3\pi$, so the relevant two-pion subsystem is naturally the isovector $P$ wave. We therefore incorporate the leading elastic rescattering effect through the complex Omnès function $\Omega_1(s+i0)$ over the physical Dalitz region, retaining both its $s$-dependent magnitude and its phase $\delta_1^1(s)$. The three two-pion channels are dressed separately by $\Omega_0=\Omega_1(s_{+-})$, $\Omega_+=\Omega_1(s_{+0})$, and $\Omega_-=\Omega_1(s_{-0})$, while the direct term is dressed by their average. This treatment is still not a complete dispersive reconstruction of $\phi\to3\pi$, because crossed-channel rescattering, a general production polynomial, and a direct bin-by-bin fit to the efficiency-corrected KLOE data are not yet included. The message of the present paper is therefore deliberately modest: the complex Omnès implementation captures an important part of the rescattering physics in a transparent amplitude-level framework, while also making clear what remains to be improved for a precision Dalitz analysis.

The paper is organized as follows. In Sec.~II, we introduce the effective interaction Lagrangians, construct the decay amplitude including the three $\rho$-exchange channels and the direct contact term, and implement the GS parametrization together with $\rho^0$-$\omega$ mixing and the complex Omnès-type FSI factor. In Sec.~III, we summarize the input parameters, present the numerical results, and compare them with the empirical decay width and the KLOE Dalitz-fit weights. Sec.~IV contains the summary and outlook.

\section{Theoretical framework}
\begin{figure}[t]
\includegraphics[width=16cm]{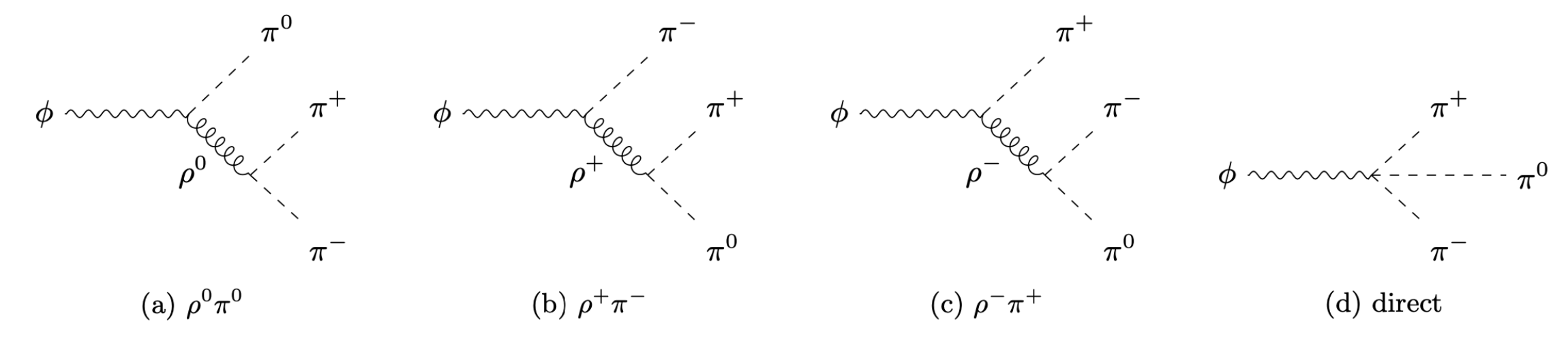}
\caption{Tree-level diagrams for $\phi\to\pi^+\pi^-\pi^0$.}
\label{fig:diag}
\end{figure}

We consider the decay process
\begin{equation}
\phi(p,\epsilon_\phi)\to\pi^+(p_+),\pi^-(p_-),\pi^0(p_0).
\end{equation}
Fig.~\ref{fig:diag} shows the tree-level contributions retained in the present framework: the three resonant $\rho$-pole channels and the direct three-pion contact term. The aim is not only to reproduce the integrated width, but also to build an amplitude-level representation in which the resonant and nonresonant pieces can be discussed separately across the full Dalitz region.

The kinematics are described by the invariant masses
\begin{equation}
s_{+-}=(p_++p_-)^2,\qquad
s_{+0}=(p_++p_0)^2,\qquad
s_{-0}=(p_-+p_0)^2,
\end{equation}
subject to
\begin{equation}
s_{+-}+s_{+0}+s_{-0}=m_\phi^2+2m_{\pi^\pm}^2+m_{\pi^0}^2.
\end{equation}
For comparison with the KLOE Dalitz analysis, we also use the variables
\begin{equation}
x=\frac{s_{-0}-s_{+0}}{2m_\phi},\qquad
y=\frac{(m_\phi-m_{\pi^0})^2-s_{+-}}{2m_\phi},
\end{equation}
which provides a one-to-one parametrization of the physical phase-space region.

The interaction Lagrangians are taken as
\begin{align}
\mathcal L_{\phi\rho\pi}
&=g_{\phi\rho\pi},
\epsilon^{\mu\nu\alpha\beta}
(\partial_\mu\phi_\nu)(\partial_\alpha\rho_\beta^a)\pi^a,
\\
\mathcal L_{\rho\pi\pi}
&=g_{\rho\pi\pi},
\epsilon^{abc}\rho_\mu^a\pi^b\partial^\mu\pi^c,
\\
\mathcal L_{\phi3\pi}
&=g_{\phi3\pi},
\epsilon^{\mu\nu\alpha\beta}
\phi_\mu\partial_\nu\pi^+\partial_\alpha\pi^-\partial_\beta\pi^0 .
\end{align}
The first two terms generate the resonant process $\phi\to\rho\pi\to\pi^+\pi^-\pi^0$ through the three charge channels $\rho^0\pi^0$, $\rho^+\pi^-$, and $\rho^-\pi^+$, whereas the last term represents the direct contribution. In the present work, the direct term is treated as an effective phenomenological parametrization of the non-$\rho\pi$ amplitude component.

In the $\phi$ rest frame, the decay amplitude can be written as
\begin{equation}
\mathcal M_{\mathrm{tot}}(\lambda)
=i m_\phi \mathcal F(s_{+-},s_{+0},s_{-0})
(\bm\epsilon^{,(\lambda)}\cdot\bm n),
\end{equation}
where the normal vector to the decay plane is $\bm n\equiv\bm p_+\times\bm p_-$. The reduced scalar amplitude is decomposed as 
\begin{equation}
\mathcal F(s_{+-},s_{+0},s_{-0})=\mathcal F_{\mathrm{dir}}+\mathcal F_{\rho\pi}+\mathcal F_{\rm BKG},
\end{equation}
with
\begin{equation}
\mathcal F_{\mathrm{dir}}=g_{\phi3\pi},\quad
\mathcal F_{\rho\pi}=-2g_{\phi\rho\pi}g_{\rho\pi\pi}\sum_{a=0,\pm}
\frac{1}{D^{\mathrm{GS}}_{\rho_a}(s_a)},\quad \mathcal F_{\rm BKG}\in {\bf R},
\end{equation}
where $s_0=s_{+-}$, $s_+=s_{+0}$, and $s_-=s_{-0}$.  Note that we also include a constant background contribution to improve the fit. This form separates the constant direct and background terms from the Dalitz-dependent resonant contribution, making their interference explicit.

After averaging over the initial $\phi$ polarizations, the squared amplitude becomes
\begin{equation}
\overline{|\mathcal M_{\mathrm{tot}}|^2}
=\frac{1}{3}m_\phi^2
|\bm p_+\times\bm p_-|^2
|\mathcal F(s_{+-},s_{+0},s_{-0})|^2,
\end{equation}
where the kinematic factor is
\begin{equation}
|\bm p_+\times\bm p_-|^2
=\frac{
\lambda(m_\phi^2,m_{\pi^\pm}^2,s_{-0})
\lambda(m_\phi^2,m_{\pi^\pm}^2,s_{+0})
}{16m_\phi^4}
-\frac{(s_{+-}-2m_{\pi^\pm}^2)^2}{4}.
\end{equation}
The double differential decay width is then
\begin{equation}
\frac{d^2\Gamma}{ds_{+-},ds_{+0}}
=\frac{1}{(2\pi)^3}
\frac{1}{32m_\phi^3}
\overline{|\mathcal M_{\mathrm{tot}}|^2}.
\end{equation}

Equations above define the tree-level kinematic normalization and the decomposition convention. In the numerical calculation, the reduced scalar amplitude $\mathcal F$ is replaced by the FSI-improved amplitude $\mathcal F^{\rm FSI}$ defined in Eq.~(\ref{eq:FFSI_full}).

No additional overall normalization factor is introduced in the final numerical convention. Instead, the off-shell production strength is absorbed into effective couplings, denoted below by $g_{\phi\rho\pi}^{\rm eff}$ and $g_{\phi3\pi}^{\rm eff}$. This convention avoids introducing a separate artificial factor whose only role would be to restore the integrated width.

To describe the broad $\rho$ resonance across the full Dalitz region, we use the GS parametrization~\cite{Gounaris:1968mw},
\begin{equation}
D^{\mathrm{GS}}_{\rho_a}(s)
=m_{\rho_a}^2-s+f_a(s)-im_{\rho_a}\Gamma_{\rho_a}(s),
\end{equation}
with the running width
\begin{equation}
\Gamma_{\rho_a}(s)
=\Gamma_{\rho_a}(m_{\rho_a}^2)
\left(\frac{m_{\rho_a}^2}{s}\right)^{1/2}
\left[\frac{k_a(s)}{k_{a\rho}}\right]^3.
\end{equation}
The GS form is particularly suitable here because the kinematically accessible two-pion invariant mass spans a wide interval, including the phase-space boundaries where a constant-width Breit-Wigner form becomes unreliable.

The neutral channel is further dressed by $\rho^0$-$\omega$ mixing according to
\begin{equation}
\frac{1}{D^{\mathrm{GS}}_{\rho^0}(s_{+-})}
\to
\frac{1}{D^{\mathrm{GS}}_{\rho^0}(s_{+-})}
\left[
1+\delta_{\rho\omega}
\frac{m_\omega^2}{D_\omega(s_{+-})}
\right],
\label{eq:mixing}
\end{equation}
where
\begin{equation}
D_\omega(s)=m_\omega^2-s-im_\omega\Gamma_\omega(s).
\end{equation}
Phenomenological analyses of the mixing self-energy suggest representative values in the range
$\Pi_{\rho\omega}=-(3400$-$4520)~\mathrm{MeV}^2$, corresponding to
$\delta_{\rho\omega}=-(1.66$-$2.20)\times10^{-3}$~\cite{McNamee:1974vb,Coon:1987kt,Bernicha:1994re,OConnell:1995nse}. In particular, the value $\Pi_{\rho\omega}=-4225~\mathrm{MeV}^2$ leads to $\delta_{\rho\omega}=-2.16\times10^{-3}$~\cite{Bernicha:1994re}. In the present work, we employ
\begin{equation}
\delta_{\rho\omega}=-2.0\times10^{-3},
\label{eq:delta-rhoomega}
\end{equation}
which lies within this phenomenologically accepted range and generates a narrow localized structure in the neutral channel on top of the broader $\rho$ background.

Although the formulation above is constructed at tree level, the dominant $\pi\pi$ final-state interaction should be incorporated if one wishes to describe the $\rho$-dominated two-pion subsystem more realistically. This extension is motivated by the structure of the decay itself: $\phi\to\pi^+\pi^-\pi^0$ proceeds mainly through $\phi\to\rho\pi\to3\pi$, so that the relevant rescattering channel is naturally the $I=1$, $P$-wave $\pi\pi$ system. The purpose of the present implementation is therefore to include the leading elastic rescattering effect in a transparent phenomenological form, while maintaining an explicit separation between the resonant and direct contributions.

The natural framework for implementing the leading elastic rescattering effect is the Omnès function for the $I=1$, $\ell=1$ channel~\cite{Omnes:1958hv,Watson:1952ji},
\begin{equation}
\Omega_1(s)
=\exp\left[
\frac{s}{\pi}
\int_{4m_\pi^2}^{\Lambda^2}ds',
\frac{\delta_1^1(s')}{s'(s'-s-i\epsilon)}
\right],
\end{equation}
where $\delta_1^1(s)$ denotes the elastic $\pi\pi$ phase shift in the isovector $P$-wave channel and $\Lambda$ is the dispersive cutoff. High-precision determinations of the $\pi\pi$ $P$-wave phase shift are available from Roy-equation analyses and related dispersive treatments~\cite{Colangelo:2001df,Garcia-Martin:2011iqs,Caprini:2005zr,Pelaez:2015qba}. In the present phenomenological study, we use the standard $\rho$-dominance phase-shift model specified below as input to the dispersion integral.

By construction,
\begin{equation}
\Omega_1(s+i0)=|\Omega_1(s)|e^{i\delta_1^1(s)}
\label{eq:omnes_phase}
\end{equation}
along the physical cut, so that the full Omnès factor carries a nontrivial $s$-dependent magnitude and phase. In particular, near $s=m_\rho^2$ one has $\delta_1^1\simeq\pi/2$, and the Omnès factor is approximately imaginary. In contrast to a constant on-shell approximation, the present calculation keeps this complex phase explicitly. This fixes the phase convention for resonant-direct interference and allows the final-state interaction to modify the Dalitz distribution in both magnitude and phase.

The elastic $P$-wave phase shift is modeled by the standard $\rho$-dominance form,
\begin{equation}
\delta_1^1(s)
=\arctan\left[
\frac{m_\rho\Gamma_\rho(s)}{m_\rho^2-s}
\right]
+\pi\theta(s-m_\rho^2),
\end{equation}
with the running width chosen consistently with the GS parametrization,
\begin{equation}
\Gamma_\rho(s)
=\Gamma_\rho(m_\rho^2)
\left(\frac{m_\rho^2}{s}\right)^{1/2}
\left[\frac{k(s)}{k_\rho}\right]^3,
\qquad
k(s)=\frac{\sqrt{\lambda(s,m_\pi^2,m_\pi^2)}}{2\sqrt{s}},
\qquad
k_\rho=k(m_\rho^2).
\end{equation}
The $\theta(s-m_\rho^2)$ correction ensures that $\delta_1^1$ rises monotonically from zero at threshold, passes through $\pi/2$ at the $\rho$ peak, and continues toward $\pi$ at high energies, consistent with the expected elastic $P$-wave behavior.

For the three charge channels, we define
\begin{equation}
\Omega_0\equiv\Omega_1(s_{+-}),\qquad
\Omega_+\equiv\Omega_1(s_{+0}),\qquad
\Omega_-\equiv\Omega_1(s_{-0}).
\end{equation}
For the nonresonant direct term, we use the channel-averaged factor
\begin{equation}
\Omega_{\rm dir}(s_{+-},s_{+0},s_{-0})
=\frac{1}{3}
\left[
\Omega_0+\Omega_++\Omega_-
\right].
\label{eq:Omega_dir}
\end{equation}
This prescription is not a substitute for a complete dispersive production polynomial, but it ensures that the direct three-pion amplitude is not artificially excluded from the same final-state interaction.

The coupling $g_{\rho\pi\pi}=5.98$ used as input is the physical coupling obtained from the empirical $\rho\to\pi\pi$ width. Since the same elastic $\pi\pi$ final-state interaction is now included explicitly through $\Omega_1(s)$, we avoid double counting by introducing
\begin{equation}
g_{\rho\pi\pi}^{\rm bare}
=\frac{g_{\rho\pi\pi}^{\rm phys}}{|\Omega_1(m_\rho^2)|}.
\label{eq:grhopipi_bare}
\end{equation}
Thus, the Omnès-dressed $\rho$-exchange amplitude is normalized consistently with the physical $\rho\to\pi\pi$ coupling at the pole. Only the modulus of $\Omega_1(m_\rho^2)$ is absorbed into $g_{\rho\pi\pi}^{\rm bare}$; the phase of the Omnès function is not discarded, but remains explicitly in the channel-dependent factors $\Omega_a(s_a)$ appearing in Eq.~(\ref{eq:FFSI_full}).

With these conventions, the resonant contribution to the reduced amplitude becomes
\begin{equation}
\mathcal F_{\rho\pi}^{\mathrm{FSI}}(s_{+-},s_{+0},s_{-0})
=-2g_{\phi\rho\pi}^{\rm eff}g_{\rho\pi\pi}^{\rm bare}
\sum_{a=0,\pm}
\frac{\Omega_a(s_a)}
{D^{\mathrm{GS}}_{\rho_a}(s_a)} .
\end{equation}
The direct term is written as
\begin{equation}
\mathcal F_{\mathrm{dir}}^{\mathrm{FSI}}
=g_{\phi3\pi}^{\rm eff}
e^{i\delta_{\rm dir}}
\Omega_{\rm dir}(s_{+-},s_{+0},s_{-0}),
\end{equation}
so that the full reduced amplitude becomes
\begin{equation}
\mathcal F^{\mathrm{FSI}}(s_{+-},s_{+0},s_{-0})
=g_{\phi3\pi}^{\rm eff}
e^{i\delta_{\rm dir}}
\Omega_{\rm dir}(s_{+-},s_{+0},s_{-0})
-2g_{\phi\rho\pi}^{\rm eff}g_{\rho\pi\pi}^{\rm bare}
\sum_{a=0,\pm}
\frac{\Omega_a(s_a)}
{D^{\mathrm{GS}}_{\rho_a}(s_a)}+\mathcal F_\mathrm{BKG}.
\label{eq:FFSI_full}
\end{equation}
This form preserves the explicit separation among the direct, resonant, and background pieces. The background term is not intended to represent an additional physical three-pion production mechanism with its own final-state rescattering. It 
is introduced only as a residual real constant accounting for smooth strength not captured by the minimal resonant-plus-direct ansatz. Hence, we do not dress it with an additional Omnès factor. The relative phase $\delta_{\rm dir}$ is defined with respect to the phase convention of the $\rho\pi$ exchange amplitude. In the numerical study below, it is treated as one of the four phenomenological fit parameters, together with $g_{\phi3\pi}^{\rm eff}$, $g_{\phi\rho\pi}^{\rm eff}$, and $\mathcal F_{\rm BKG}$.

For the neutral channel, the $\rho^0$-$\omega$ mixing correction is retained as
\begin{equation}
\frac{\Omega_0}{D^{\mathrm{GS}}_{\rho^0}(s_{+-})}
\to
\frac{\Omega_0}{D^{\mathrm{GS}}_{\rho^0}(s_{+-})}
\left[
1+\delta_{\rho\omega}
\frac{m_\omega^2}{D_\omega(s_{+-})}
\right],
\label{eq:mixing_fsi}
\end{equation}
while the charged channels contain
\begin{equation}
\frac{\Omega_\pm}{D^{\mathrm{GS}}_{\rho^\pm}(s_{\pm0})}.
\end{equation}
Equation~(\ref{eq:FFSI_full}) is the amplitude used for the numerical results. It gives the resonant and direct mechanisms the appropriate final-state interaction dressing while retaining the background term as an additional real-constant contribution. This convention preserves the amplitude-level separation among the resonant, direct, and background components.

To focus on the intrinsic dynamical pattern, we define the reduced Dalitz density when the FSI-improved amplitude is used,
\begin{equation}
\rho_{\mathrm{red}}^{\mathrm{FSI}}(s_{+-},s_{+0})
\equiv
|\mathcal F^{\mathrm{FSI}}(s_{+-},s_{+0},s_{-0})|^2.
\end{equation}
Unlike the full Dalitz density, the reduced quantity factors out the universal phase-space geometry, making the resonance bands, the mixing-induced feature, and the direct-term interference pattern easier to identify.

Finally, because the amplitude is proportional to the normal vector of the decay plane, the polarization structure is highly constrained. In a basis where the longitudinal axis is chosen along $\hat n\equiv\bm n/|\bm n|$, the amplitude is purely longitudinal, whereas in a basis tied to an intermediate momentum, the longitudinal component vanishes and only the transverse component normal to the decay plane survives. This geometric structure remains unchanged when the amplitude is improved from $\mathcal F$ to $\mathcal F^{\mathrm{FSI}}$, because the rescattering correction modifies only the reduced scalar function and not the tensor structure of the decay amplitude. Detailed explanations of the polarizations are given in Appendix A.

\section{Numerical results and discussion}
We now summarize the numerical inputs and the main results. The coupling $g_{\rho\pi\pi}^{\rm phys}=5.98$ is fixed from the empirical $\rho\to\pi\pi$ width. Since the elastic $P$-wave $\pi\pi$ final-state interaction is included explicitly through the complex Omnès function, the $\rho$-exchange amplitude uses the bare coupling $g_{\rho\pi\pi}^{\rm bare}=g_{\rho\pi\pi}^{\rm phys}/|\Omega_1(m_\rho^2)|$. Numerically, this gives $g_{\rho\pi\pi}^{\rm bare}=1.340$--$1.341$. Here \textit{bare} does not denote a fundamental Lagrangian bare coupling, but an effective undressed coupling in the present Omnès-improved phenomenological convention. The production strengths are represented by the effective couplings $g_{\phi\rho\pi}^{\rm eff}$ and $g_{\phi3\pi}^{\rm eff}$, the relative phase of the direct amplitude is denoted by $\delta_{\rm dir}$, and the real constant background amplitude is denoted by $\mathcal{F}_\mathrm{BKG}$. In the final numerical setup used below, we take
\begin{equation}
g_{\phi\rho\pi}^{\rm eff}=6.27~{\rm GeV}^{-1},\qquad
g_{\phi3\pi}^{\rm eff}=-24.0~{\rm GeV}^{-3},\qquad
\delta_{\rm dir}=-25^\circ,
\end{equation}
together with
\begin{equation}
\Lambda_\Omega=1.60~{\rm GeV},\qquad
\mathcal{F}_\mathrm{BKG}=-50.0~{\rm GeV}^{-3},\qquad
\delta_{\rho\omega}=-0.002 .
\end{equation}
The coupling $g_{\phi\rho\pi}^{\rm eff}$ requires a special comment. In the present work, it is not extracted from a narrow-width on-shell two-body decay formula for $\phi\to\rho\pi$. Although the nominal $\rho\pi$ threshold lies below the $\phi$ mass, the physical process considered here is the three-body decay $\phi\to\pi^+\pi^-\pi^0$, where the intermediate $\rho$ is broad, is described by the running-width GS propagator, and is sampled over the full Dalitz region. Therefore, a unique on-shell two-body extraction of $g_{\phi\rho\pi}$ would not provide the coupling appropriate for the present off-shell three-body amplitude. We instead regard $g_{\phi\rho\pi}^{\rm eff}$ as a phenomenological off-shell production strength.

The effective parameters should be regarded as phenomenological quantities within the present minimal FSI-improved framework, rather than as unique results of a full Dalitz-bin fit. They are chosen by minimizing a simple diagnostic objective, not by performing a full statistical fit. Schematically, we write this objective as
\begin{equation}
\chi_{\rm diag}^2 =
w_\Gamma
\left(
\frac{\Gamma_{\rm th}-\Gamma_{\rm exp}}{\Gamma_{\rm exp}}
\right)^2
+
w_\rho
\left(
\frac{I_{\rho\pi}^{\rm th}-I_{\rho\pi}^{\rm KLOE}}
{I_{\rho\pi}^{\rm KLOE}}
\right)^2
+
w_{\rm dir}
\left(
\frac{I_{\rm dir}^{\rm th}-I_{\rm dir}^{\rm KLOE}}
{I_{\rm dir}^{\rm KLOE}}
\right)^2
+
w_x\left(\chi_x^{\rm shape}\right)^2
+
w_y\left(\chi_y^{\rm shape}\right)^2.
\end{equation}
Here $\chi_x^{\rm shape}$ and $\chi_y^{\rm shape}$ denote the normalized root-mean-square differences between the theoretical and experimental one-dimensional projections after each projection has been normalized to unit area. The weights $w_i$ are used only to select a representative phenomenological parameter set and are not interpreted as elements of, or substitutes for, an experimental covariance matrix. Thus, this procedure is not a substitute for a bin-by-bin covariance fit, but simply specifies how the effective couplings used in the present comparison are fixed.

The present analysis is intentionally limited. It compares the calculated integrated quantities and one-dimensional projections with empirical information, but it does not yet use the efficiency-corrected KLOE Dalitz-bin data or their covariance matrix in a direct amplitude fit. The numerical results should therefore be viewed as diagnostic of the present dynamical setup. The input parameters, phenomenological reference values, numerical outputs, and comparison with the KLOE Dalitz-fit weights are collected in Tables~\ref{tab:input_parameters} and \ref{tab:main_results}. The empirical partial width is estimated from~\cite{ParticleDataGroup:2024cfk}
\begin{equation}
\Gamma(\phi\to\pi^+\pi^-\pi^0)=\mathrm{Br}(\phi\to\pi^+\pi^-\pi^0)\,\Gamma_\phi,
\end{equation}
using $\Gamma_\phi=4.43\pm0.05~\mathrm{MeV}$ and $\mathrm{Br}(\phi\to\pi^+\pi^-\pi^0)=(14.9\pm0.4)\%$, which gives
\begin{equation}
\Gamma_{\mathrm{exp}}(\phi\to\pi^+\pi^-\pi^0)\approx0.660\pm0.020~\mathrm{MeV}.
\end{equation}

\begin{table}[b]
\centering
\begin{ruledtabular}
\begin{tabular}{lll}
Quantity & Value & Comment \\
\hline
$g_{\rho\pi\pi}^{\rm phys}$ 
& $5.98$ 
& fixed from the physical $\rho\to\pi\pi$ width \\
$g_{\rho\pi\pi}^{\rm bare}$ 
& $1.340$--$1.341$ 
& used inside the Omn\`es-dressed $\rho$-exchange amplitude \\
$g_{\phi\rho\pi}^{\rm eff}$ 
& $6.27~{\rm GeV}^{-1}$ 
& phenomenological off-shell resonant production strength \\
$g_{\phi3\pi}^{\rm eff}$ 
& $-24.0~{\rm GeV}^{-3}$ 
& phenomenological direct production strength \\
$\delta_{\rm dir}$ 
& $-25^\circ$ 
& relative phase of the direct amplitude \\
$\delta_{\rho\omega}$ 
& $-0.002$ 
& within the phenomenological literature range \\
$\Lambda_{\Omega}$ 
& $1.60~{\rm GeV}$ 
& Omn\`es dispersive cutoff \\
$\mathcal{F}_\mathrm{BKG}$ 
& $-50.0~{\rm GeV}^{-3}$ 
& real constant background amplitude, no FSI \\
$\Gamma_\phi$ 
& $4.43\pm0.05~{\rm MeV}$ 
& PDG input \\
${\rm Br}(\phi\to\pi^+\pi^-\pi^0)$ 
& $(14.9\pm0.4)\%$ 
& PDG input \\
$\Gamma_{\rm exp}$ 
& $0.660\pm0.020~{\rm MeV}$ 
& from $\Gamma_\phi\times{\rm Br}$ \\
\end{tabular}
\end{ruledtabular}
\caption{Input parameters and phenomenological effective parameters used in the final comparison. The background term is added as a constant amplitude to the reduced amplitude, while the direct term is dressed by the averaged Omnès final-state interaction.}
\label{tab:input_parameters}
\end{table}

\begin{table}[b]
\centering
\begin{ruledtabular}
\begin{tabular}{lll}
Quantity & Present work & Comparison or comment \\
\hline
$\Gamma_{\rm th}$ 
& $0.6915~{\rm MeV}$ 
& slightly above $\Gamma_{\rm exp}$ \\
$\Gamma_{\rm th}^{(\cos\theta)}$ 
& $0.6941~{\rm MeV}$ 
& projection cross-check \\
$I_{\rho\pi}$ 
& $0.9619$ 
& KLOE: $0.937$ \\
$I_{\rm dir}$ 
& $8.393\times 10^{-3}$ 
& KLOE: $8.5\times 10^{-3}$ \\
$\sqrt{I_{\rm dir}/I_{\rho\pi}}$ 
& $9.341\times 10^{-2}$ 
& small direct component \\
$\cos\theta_{+-}^{\rm peak}$ 
& $0.9392$ 
& opening-angle peak \\
$x_{\rm peak}$ 
& $\simeq 0$ 
& $x$ projection peak \\
$y_{\rm peak}$ 
& $0.1460~{\rm GeV}$ 
& $y$ projection peak \\
$\chi_x^{\rm shape}$ 
& $0.0985$ 
& area-normalized $x$-projection NRMSE \\
$\chi_y^{\rm shape}$ 
& $0.2803$ 
& area-normalized $y$-projection NRMSE \\
$I_{\omega\pi}$ 
& --- 
& KLOE: $2.0\times 10^{-4}$; see text \\
\end{tabular}
\end{ruledtabular}
\caption{Main numerical results of the present calculation and comparison with the KLOE Dalitz-fit weights. The projection-shape errors are computed after independently normalizing the experimental and theoretical projections to unit area.}
\label{tab:main_results}
\end{table}

The present calculation gives
\begin{equation}
\Gamma_{\mathrm{th}}(\phi\to\pi^+\pi^-\pi^0)=0.6915~\mathrm{MeV},
\end{equation}
which is slightly above the empirical central value. The difference is about $4.8\%$, comparable to the phenomenological uncertainty expected from the restricted treatment of the amplitude and from the absence of a direct Dalitz-bin covariance fit. As a numerical cross-check, the angular-projection integration gives
\begin{equation}
\Gamma_{\rm th}^{(\cos\theta)}=0.6941~{\rm MeV}.
\end{equation}
The small difference between the direct Dalitz integration and this projection check reflects the finite binning and projection procedure used in the numerical integration.

To quantify the relative size of the resonant and direct pieces, we define
\begin{equation}
I_x=\frac{\int dx\,dy\,|\mathcal F_x^{\rm FSI}|^2}{\int dx\,dy\,|\mathcal F^{\rm FSI}|^2},
\qquad x=\rho\pi,\,\mathrm{dir},
\end{equation}
where $\mathcal F^{\rm FSI}$ is given in Eq.~(\ref{eq:FFSI_full}). Using the same normalization convention as in the KLOE Dalitz-fit analysis, we obtain
\begin{equation}
I_{\rho\pi}=0.9619,\qquad I_{\mathrm{dir}}=8.393\times10^{-3}.
\end{equation}
The corresponding direct-to-resonant amplitude-scale ratio is
\begin{equation}
\sqrt{\frac{I_{\rm dir}}{I_{\rho\pi}}}=9.341\times10^{-2}.
\end{equation}
Compared with the KLOE values $I_{\rho\pi}=0.937$ and $I_{\mathrm{dir}}=8.5\times10^{-3}$~\cite{KLOE:2003kas}, the direct component is reproduced at essentially the same level, whereas the resonant weight comes out slightly larger in the present parametrization.

This pattern should not be interpreted as a precision determination of the resonant fraction. In the present formulation, the full complex Omnès factor is retained, but the FSI treatment is still limited to a minimal single-channel elastic implementation. Crossed-channel rescattering, a general dispersive production polynomial, an explicit independent $\omega\pi^0$ amplitude, and a direct fit to the efficiency-corrected KLOE Dalitz bins are not yet included. The comparison with KLOE is therefore informative mainly in two respects: it confirms that the direct sector remains at the correct scale, and it shows that the detailed redistribution of resonant strength requires a more complete dispersive and data-level treatment.

A related caution applies to the often-quoted $\sim6\%$ resonant-direct interference extracted by KLOE~\cite{KLOE:2003kas}. The present framework is consistent with the qualitative importance of that interference, but it does not yet determine its detailed energy dependence with the accuracy of a full dispersive amplitude fit. Accordingly, the present value of $I_{\rho\pi}$ should be read as a useful phenomenological indicator, not as a final determination.

Regarding the $\omega\pi^0$ component reported by KLOE ($I_{\omega\pi}=2.0\times10^{-4}$), the present model does not include a separate $\omega\pi^0$ amplitude in the sense of the KLOE three-component decomposition. Nevertheless, the $\rho^0$-$\omega$ mixing prescription of Eq.~(\ref{eq:mixing}) implicitly generates a narrow $\omega$-like enhancement in the neutral two-pion channel through the $\delta_{\rho\omega}m_\omega^2/D_\omega(s_{+-})$ correction to the $\rho^0$ propagator. The resulting localized deformation visible in the neutral band of the Dalitz plot is physically the same effect as the $\omega\pi^0$ contribution identified by KLOE, even though it enters here via the mixed propagator rather than as an independent amplitude. Since $I_{\omega\pi} = 2.0\times10^{-4}$ is two orders of magnitude smaller than $I_{\mathrm{dir}}$, its omission from the present fit is not expected to affect the determination of $g_{\phi3\pi}^{\rm eff}$, $g_{\phi\rho\pi}^{\rm eff}$, or $\delta_{\rm dir}$ at the level of precision targeted here. A direct numerical comparison with $I_{\omega\pi}$ would require isolating the $\omega$-pole residue from the full mixed propagator and is left for future work.

\begin{figure}[t]
\topinset{(a)}{\includegraphics[width=6.5cm]{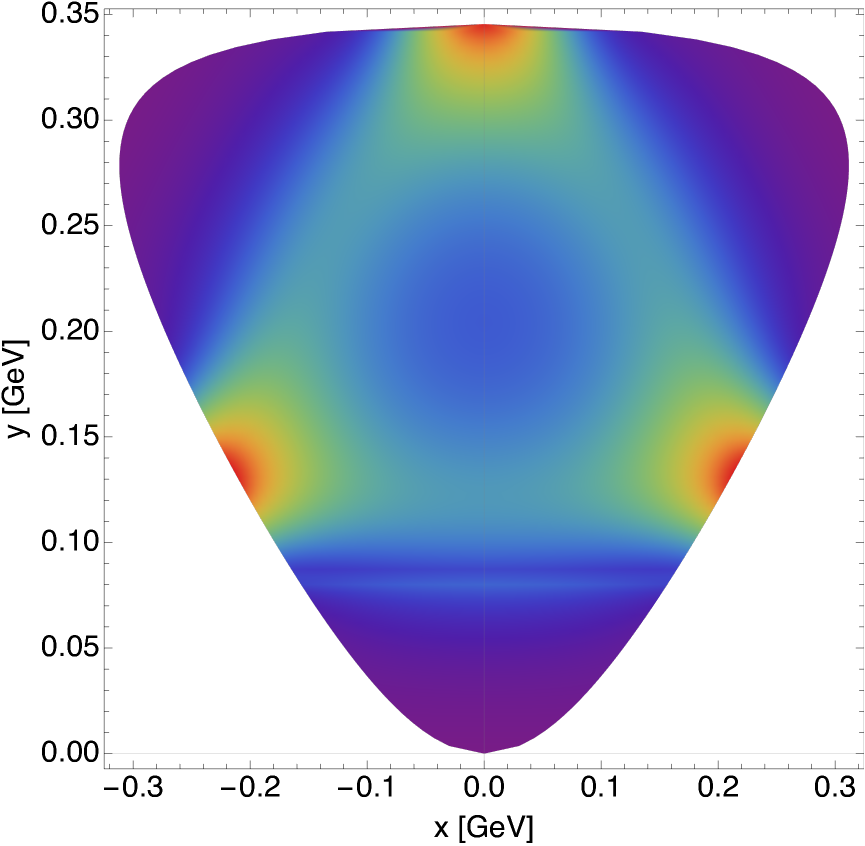}}{-0.4cm}{0.0cm}
\hspace{1cm}
\topinset{(b)}{\includegraphics[width=7cm]{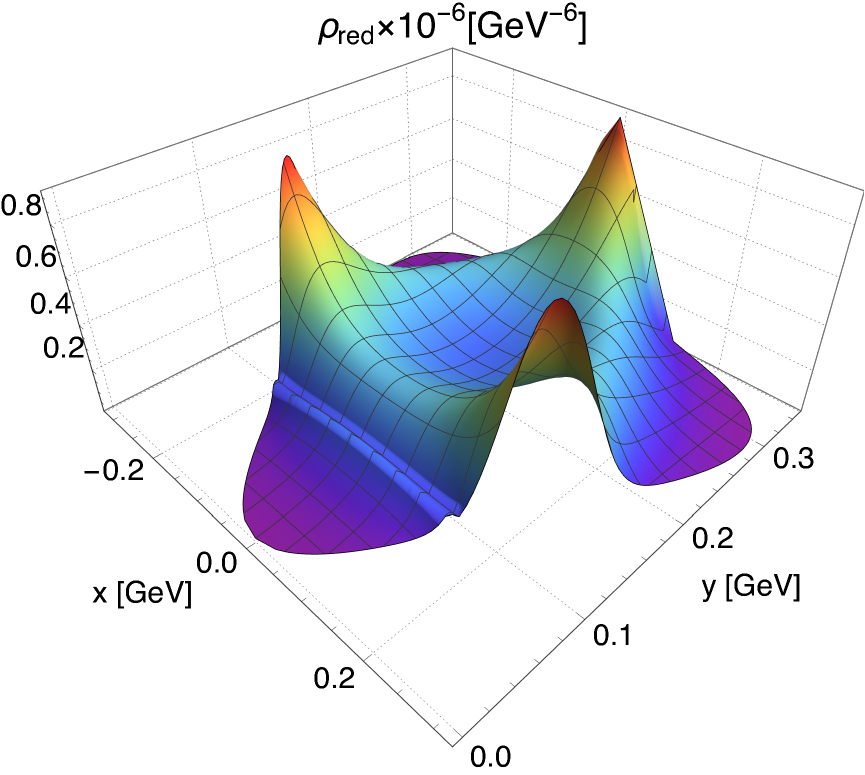}}{-0.4cm}{0.0cm}
\caption{(Color online) Reduced Dalitz-plot density with the FSI for $\phi\to\pi^+\pi^-\pi^0$. Panel (a) shows the two-dimensional distribution in the $(x,y)$ plane, while panel (b) displays the corresponding three-dimensional representation. The dominant three-band structure arises from the $\rho^0\pi^0$, $\rho^+\pi^-$, and $\rho^-\pi^+$ intermediate states, and the neutral channel exhibits a localized deformation associated with the $\rho^0$-$\omega$ mixing term.}
\label{fig:dalitz}
\end{figure}

Fig.~\ref{fig:dalitz} shows the reduced Dalitz-plot density for $\phi\to\pi^+\pi^-\pi^0$ obtained with the FSI-improved amplitude of Eq.~(\ref{eq:FFSI_full}). In the $(x,y)$ variables, one clearly sees the dominant three-band pattern generated by the $\rho^0\pi^0$, $\rho^+\pi^-$, and $\rho^-\pi^+$ intermediate states. A narrow localized structure is also visible in the neutral channel. Since the mixing correction enters through the neutral two-pion invariant mass, this localized feature is naturally identified with the $\rho^0$-$\omega$ term. The global Dalitz profile is governed by the resonant $\rho$ exchange, while the mixing term produces only a localized deformation.

A central point of the present formulation is that the direct term does not generate an isolated resonance-like band of its own. Instead, it modifies the Dalitz density mainly through interference with the dominant resonant amplitude. This is precisely why a direct three-body contribution is difficult to identify from the raw density alone and why an amplitude-level separation is useful. In the present setup, the direct component appears as a smooth distortion superimposed on the much larger three-band resonant pattern.

\begin{figure}[t]
\topinset{(a)}{\includegraphics[width=7.5cm]{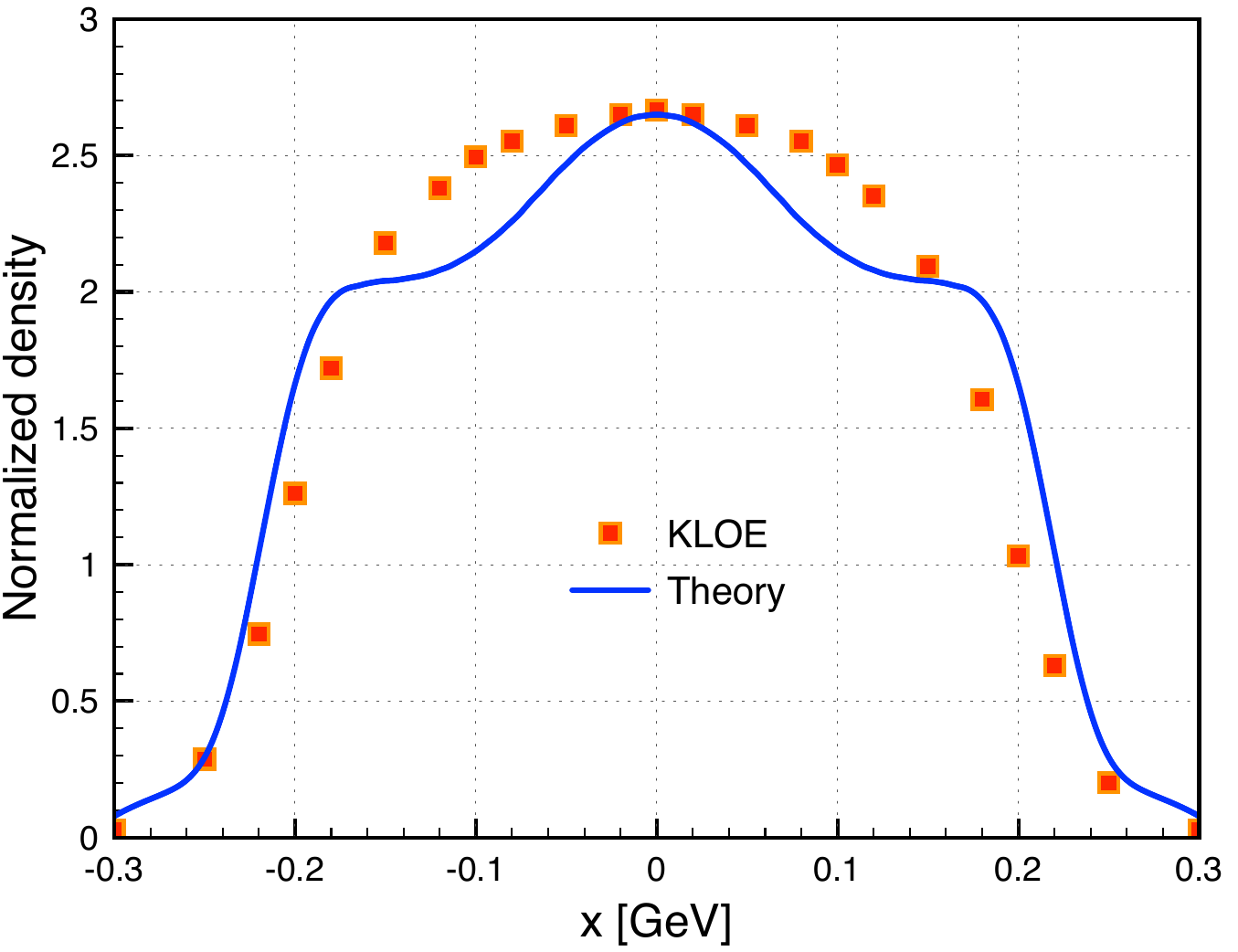}}{-0.4cm}{0.0cm}
\topinset{(b)}{\includegraphics[width=7.5cm]{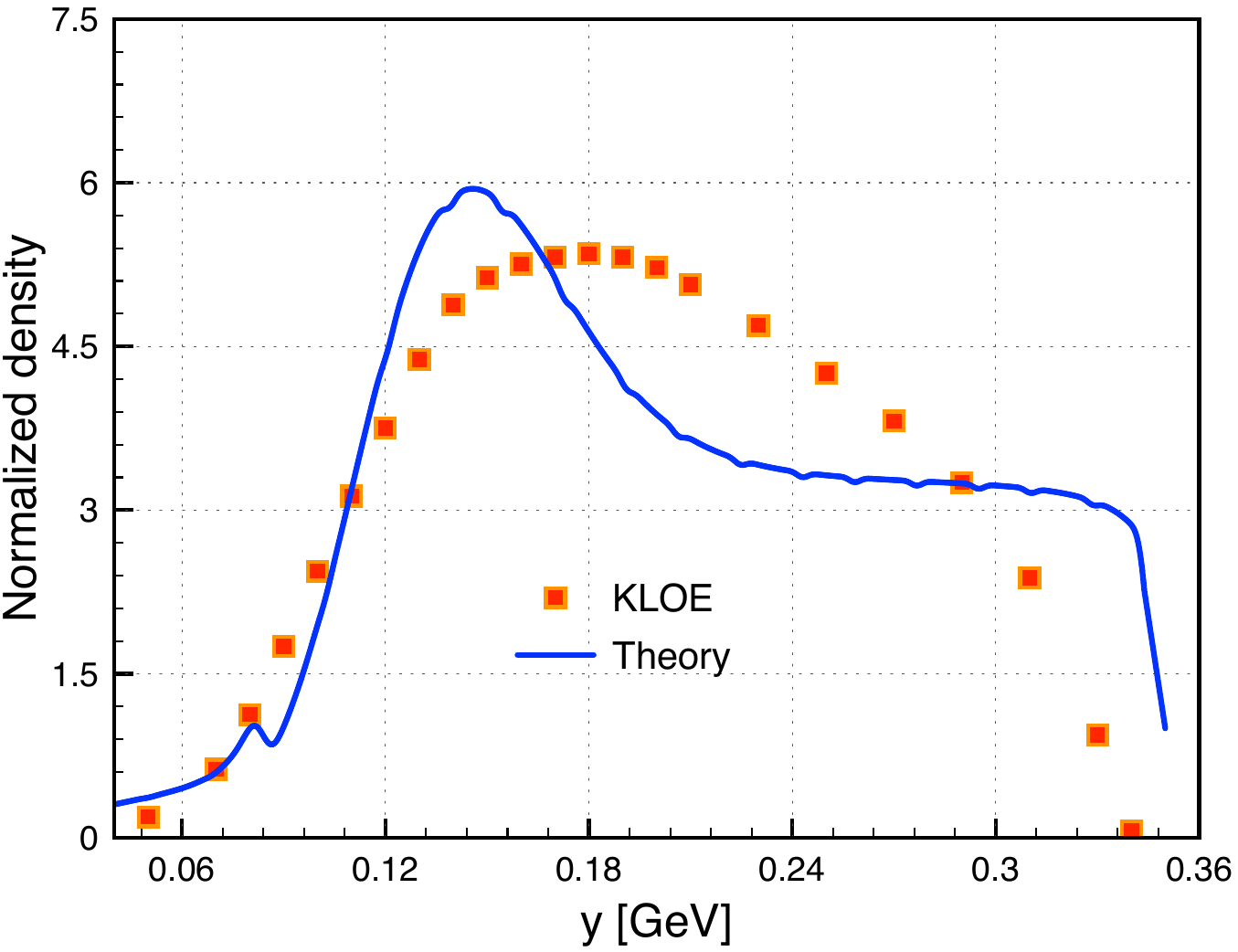}}{-0.4cm}{0.0cm}
\caption{(Color online) Normalized $x$ and $y$ projections of the reduced Dalitz-plot density obtained with the complex Omnès-type FSI, compared with the KLOE data. The cumulative integrals are also shown to compare the normalization trends.}
\label{fig:proj}
\end{figure}

Fig.~\ref{fig:proj}(a) and (b) show the one-dimensional $x$ [GeV] and $y$ [GeV] projections of the reduced Dalitz densities, respectively, with the KLOE data, along with the corresponding cumulative integrals for normalization comparison. The $x$ projection is broad and approximately symmetric, with its peak located near
\begin{equation}
x_{\rm peak}\simeq 0 .
\end{equation}
This reflects the near balance between the $\rho^+\pi^-$ and $\rho^-\pi^+$ channels under interchange of the two charged pions. By contrast, the $y$ projection is more structured, with a peak near
\begin{equation}
y_{\mathrm{peak}}=0.1460~\mathrm{GeV}
\end{equation}
in the present calculation. This difference is natural because $y$ is tied directly to $s_{+-}$ and therefore probes the neutral channel, including the localized $\rho^0$-$\omega$ effect, more directly.

The area-normalized projection-shape deviations are quantified by
\begin{equation}
\chi_x^{\rm shape}=0.0985,\qquad
\chi_y^{\rm shape}=0.2803 .
\end{equation}
Thus, the $x$ projection is reproduced at a reasonable level, whereas the $y$ projection still shows a more pronounced discrepancy. A more detailed examination reveals characteristic shape mismatches in both projections. In the $x$ projection, the theoretical curve remains broader than the KLOE distribution: the tails at both $x\lesssim-0.15$ and $x\gtrsim+0.15$ are overestimated, indicating that the present framework spreads too much strength into the peripheral phase-space region. In the $y$ projection, the theoretical curve overestimates part of the high-$y$ tail and does not fully reproduce the sharp behavior near the kinematic boundary. These systematic features indicate that the current minimal FSI-improved amplitude remains insufficient to capture the detailed shape of the Dalitz projections near the phase-space boundary.

This residual discrepancy is physically understandable. First, the large-$y$ region lies close to the kinematical edge of the Dalitz domain, where phase-space suppression and boundary geometry become increasingly important. Second, although the complex Omnès-type correction included here accounts for the leading elastic $P$-wave $\pi\pi$ rescattering effect, the treatment remains a minimal single-channel implementation rather than a complete dispersive representation that includes crossed-channel rescattering. Third, the direct contribution is modeled with an effective constant production strength dressed by the averaged Omnès factor, which is useful for tracking its interference with the dominant resonant amplitude but may be too restrictive to reproduce the detailed energy dependence of the projected spectra. Fourth, the present comparison is not yet based on a direct fit to the efficiency-corrected Dalitz-bin data with the full covariance information used in the KLOE analysis, where the detector efficiency varies over the Dalitz plane and the small $\omega\pi^0$ structure was included explicitly~\cite{KLOE:2003kas}. For these reasons, the current framework should be regarded as a quantitatively improved but still incomplete description of the projection-level observables, especially near the phase-space boundary.

The implications for future work are therefore clear. A next-stage analysis should fit the model directly to the KLOE Dalitz-bin data using the experimental binning, efficiency map, and covariance information rather than relying only on projection-level comparisons. The present Omnès-based FSI treatment should also be extended to a more complete dispersive framework, allowing for crossed-channel rescattering and the systematic incorporation of a more realistic analytic structure. In addition, the direct term should be generalized beyond the present constant effective coupling to allow for mild energy dependence, and the $\omega\pi^0$ contribution should be included as a separately identifiable amplitude in the same fitting framework. These steps are necessary if one aims to describe not only the integrated width and the gross Dalitz pattern, but also the detailed shapes of the $x$ and $y$ projections over the full physical region.

\begin{figure}[t]
\includegraphics[width=7.5cm]{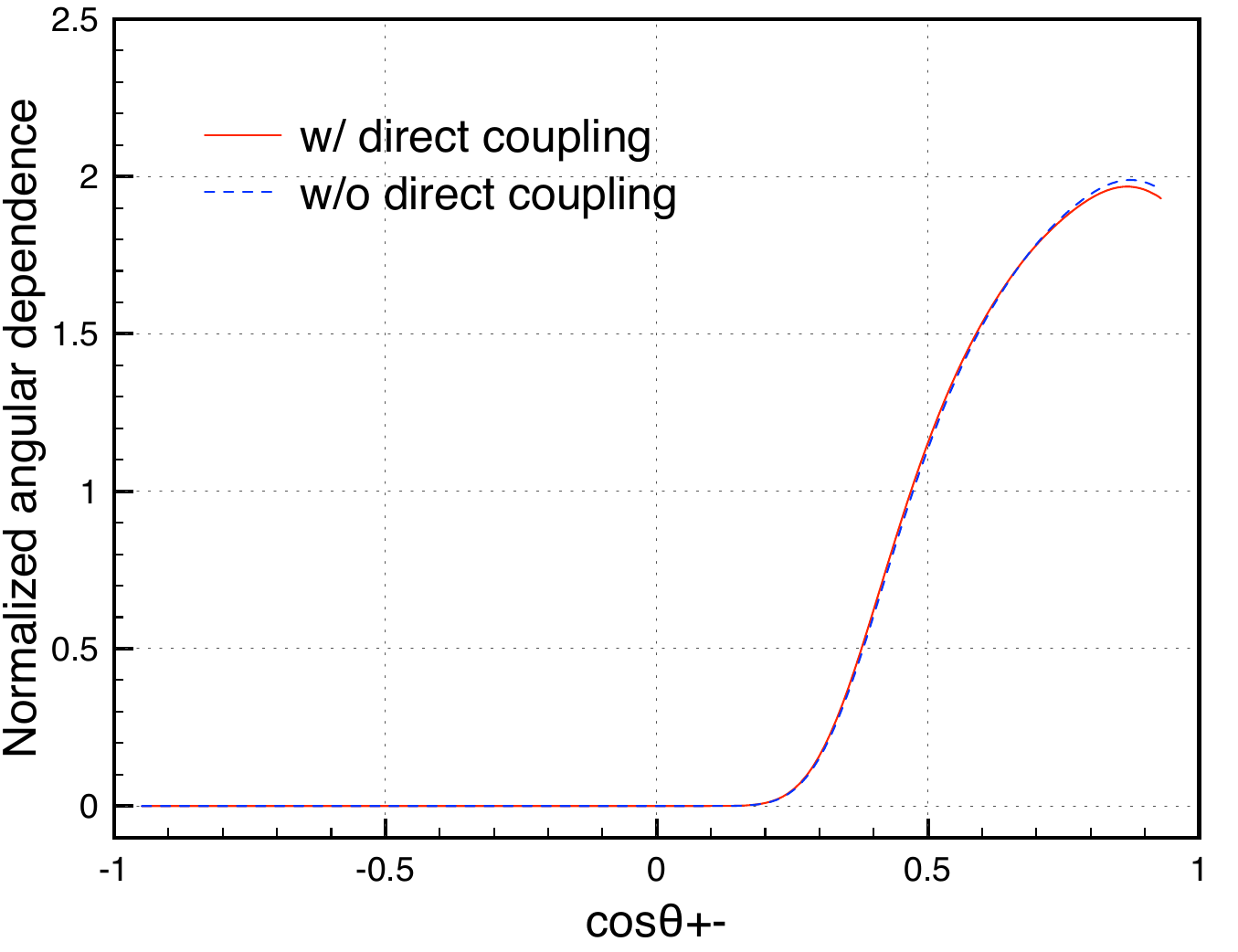}
\caption{(Color online) Normalized angular distribution $(1/\Gamma)d\Gamma/d\cos\theta_{+-}$ with the FSI as a function of $\cos\theta_{+-}$. The solid and dashed lines correspond to the results with and without the direct coupling, respectively.}
\label{fig:ang}
\end{figure}

Fig.~\ref{fig:ang} shows the normalized angular distribution, i.e., $(1/\Gamma)d\Gamma/d\cos\theta_{+-}$,  with respect to the relative opening angle between the charged pions. The definition of the angular dependence is explained in detail in Appendix B. The distribution is strongly nonuniform and peaks at large positive $\cos\theta_{+-}$, with the maximum located at
\begin{equation}
\cos\theta_{+-}^{\mathrm{peak}}=0.9392.
\end{equation}
This indicates that the kinematically weighted decay region favors relatively small opening angles between the charged pions. The angular observable does not introduce an independent dynamical ingredient; rather, it is another projection of the same underlying Dalitz dynamics, governed jointly by the phase-space factor and the resonant structure of the reduced amplitude.

Several observables were considered as potential handles for isolating the direct contribution $g_{\phi3\pi}^{\rm eff}$ from the dominant resonant amplitude. The opening-angle distribution shown in Fig.~\ref{fig:ang}, the azimuthal interference pattern analogous to a generalized parton distribution (GPD) decomposition, and the asymmetry of the Dalitz density under interchange of the two charged pions were all examined in this context, but each was found to be insufficient for this purpose in the present setup. The opening-angle distribution is strongly peaked at $\cos\theta_{+-} \approx 1$ regardless of whether the direct term is included: as seen in Fig.~\ref{fig:ang}, the two curves corresponding to $g_{\phi3\pi}^{\rm eff}=-24.0\,\mathrm{GeV}^{-3}$ with $\delta_{\rm dir}=-25^\circ$ and to $g_{\phi3\pi}^{\rm eff}=0$ differ by less than $1\%$ over the full range of $\cos\theta_{+-}$, well below any practically accessible statistical resolution. This insensitivity is a direct consequence of the smallness of the direct component ($I_{\mathrm{dir}}/I_{\rho\pi} \sim 10^{-2}$): the direct term contributes at a level well below the statistical resolution of any practically achievable modification to this one-dimensional projection.

The azimuthal interference between different helicity components, which in principle could provide an additional handle via an angular asymmetry in the decay plane, is also unavailable here. Because the full amplitude is proportional to the normal vector $\mathbf{n} \equiv \mathbf{p}_{+} \times \mathbf{p}_{-}$, all angular information is already encoded in $\mathbf{n}$, and no independent azimuthal degree of freedom survives after the polarization sum; the azimuthal structure collapses entirely into the Dalitz dynamics. Finally, the left-right asymmetry of the Dalitz density under $s_{+0} \leftrightarrow s_{-0}$, which reflects the interference between the $\rho^{+}\pi^{-}$ and $\rho^{-}\pi^{+}$ channels modulated by the direct term, is in principle nonzero when $g_{\phi3\pi} \neq 0$, but its magnitude is too small to be discriminated at the level of the present framework. These considerations collectively suggest that isolating the direct contribution requires an amplitude-level fit directly to the efficiency-corrected Dalitz-bin data, rather than any single projected observable.

The overall numerical picture may therefore be summarized as follows. The global Dalitz pattern is dominated by the $\rho\pi$ mechanism; the direct term remains small but phenomenologically relevant through interference; and the neutral channel carries a localized deformation from $\rho^0$-$\omega$ mixing. Within this framework, the complex Omnès factor provides a nontrivial final-state-interaction dressing of both the resonant and direct mechanisms. At the same time, the present results also make explicit their own limit of validity. The remaining tension in the projected Dalitz distributions indicates that the minimal single-channel FSI implementation and the restricted fit strategy are insufficient to precisely describe the full Dalitz structure. The present framework should therefore be understood as an intermediate phenomenological step: it captures an important rescattering effect in a transparent form, but it does not eliminate the need for a complete dispersive treatment and a data-level fit.

\section{Summary and outlook}
We have analyzed the three-body decay $\phi\to\pi^+\pi^-\pi^0$ in an effective-Lagrangian framework that explicitly separates the dominant resonant $\rho\pi$ mechanism, the direct three-pion term, and a real constant background contribution at the amplitude level. The resonant contribution is described with the Gounaris-Sakurai parametrization, the neutral channel includes $\rho^0$-$\omega$ mixing, and the leading elastic $I=1$, $P$-wave $\pi\pi$ rescattering effect is incorporated through the complex, $s$-dependent Omnès function $\Omega_1(s+i0)$. The three two-pion channels are dressed separately by $\Omega_0$, $\Omega_+$, and $\Omega_-$, while the direct term is multiplied by the averaged factor $\Omega_{\rm dir}=(\Omega_0+\Omega_++\Omega_-)/3$. The background contribution is retained as a real constant amplitude and is not dressed by the Omnès final-state interaction.

A technical point of the revised formulation is the separation between the physical $\rho\pi\pi$ coupling and the explicit final-state-interaction factor. Since $g_{\rho\pi\pi}^{\rm phys}=5.98$ already contains the empirical strength of $\rho\to\pi\pi$, the Omnès-dressed exchange amplitude uses $g_{\rho\pi\pi}^{\rm bare}=g_{\rho\pi\pi}^{\rm phys}/|\Omega_1(m_\rho^2)|$ to avoid double counting. Numerically, this gives $g_{\rho\pi\pi}^{\rm bare}=1.340$--$1.341$. The remaining production strengths are absorbed into the effective couplings $g_{\phi\rho\pi}^{\rm eff}=6.27~{\rm GeV}^{-1}$ and $g_{\phi3\pi}^{\rm eff}=-24.0~{\rm GeV}^{-3}$, together with the relative direct phase $\delta_{\rm dir}=-25^\circ$ and the constant background amplitude $\mathcal{F}_\mathrm{BKG}=-50.0~{\rm GeV}^{-3}$. These quantities are treated as phenomenological off-shell parameters and are fixed by the diagnostic objective described in Sec.~III, not by an on-shell two-body $\phi\to\rho\pi$ width extraction.

With the final parameter set used in the present analysis, the calculated width is $\Gamma_{\rm th}(\phi\to\pi^+\pi^-\pi^0)=0.6915~{\rm MeV}$, which is slightly above the empirical estimate $\Gamma_{\rm exp}\approx0.660\pm0.020~{\rm MeV}$. The angular-projection cross-check gives $\Gamma_{\rm th}^{(\cos\theta)}=0.6941~{\rm MeV}$. The direct integrated weight is $I_{\rm dir}=8.393\times10^{-3}$, which remains essentially at the KLOE scale, while the resonant weight is $I_{\rho\pi}=0.9619$. The physical picture that emerges is clear: the broad three-band structure of the Dalitz plot is controlled by the $\rho\pi$ mechanism, the direct term remains small but relevant through interference, and the neutral channel carries a localized $\rho^0$-$\omega$ deformation.

The present treatment does not yet provide a precise description of the full Dalitz surface. The remaining broad tails in the $x$ projection and the mismatch in the $y$ distribution near the phase-space boundary should not be attributed to a constant on-shell Omnès approximation, because the revised calculation already keeps the complex $s$-dependent Omnès factor. Instead, these residual discrepancies reflect the limitations of the present minimal single-channel elastic FSI implementation, the use of an effective constant direct production strength, the absence of crossed-channel rescattering, the lack of an explicitly separated $\omega\pi^0$ amplitude, and the fact that the calculation has not yet been fitted directly to the efficiency-corrected KLOE Dalitz-bin data with their covariance matrix.

The next steps are therefore well defined. The present Omnès-based amplitude should be embedded in a complete dispersive framework including crossed-channel rescattering and the proper analytic structure, along the lines of Refs.~\cite{Niecknig:2012sj,Danilkin:2014cra}. A direct fit to the KLOE Dalitz-bin data should then be performed using the experimental binning, efficiency map, and covariance information. In the same framework, the direct term should be generalized to an energy-dependent production polynomial, and the $\omega\pi^0$ contribution should be isolated as an explicit amplitude component. These steps are required for a quantitatively controlled description of the full $\phi\to\pi^+\pi^-\pi^0$ Dalitz structure. Related works will appear elsewhere.

\section*{Acknowledgment}
This work was supported by grants from the National Research Foundation of Korea (NRF), funded by the Korean government (MSIT) (NRF-RS-2025-16065906 and NRF-RS-2024-00436392).
\appendix
\section{Polarization decomposition of the $\phi$ meson}
For the polarization decomposition, we introduce an orthonormal basis in the $\phi$-meson rest frame, $\hat{\mathbf e}_L$, $\hat{\mathbf e}_{T1}$, and $\hat{\mathbf e}_{T2}$, and define
\begin{equation}
\epsilon_L^\mu=(0,\hat{\mathbf e}_L),\qquad
\epsilon_{T1}^\mu=(0,\hat{\mathbf e}_{T1}),\qquad
\epsilon_{T2}^\mu=(0,\hat{\mathbf e}_{T2}).
\end{equation}
The corresponding amplitudes are
\begin{equation}
\mathcal M_L=i\,m_\phi\,\mathcal F\,(\hat{\mathbf e}_L\cdot\mathbf n),\qquad
\mathcal M_{T1}=i\,m_\phi\,\mathcal F\,(\hat{\mathbf e}_{T1}\cdot\mathbf n),\qquad
\mathcal M_{T2}=i\,m_\phi\,\mathcal F\,(\hat{\mathbf e}_{T2}\cdot\mathbf n),
\end{equation}
where
\begin{equation}
\mathbf n\equiv\mathbf p_+\times\mathbf p_-.
\end{equation}
Choosing $\hat{\mathbf e}_L=\hat{\mathbf n}\equiv\mathbf n/|\mathbf n|$, one finds
\begin{equation}
\mathcal M_L=i\,m_\phi\,|\mathbf n|\,\mathcal F,\qquad
\mathcal M_{T1}=0,\qquad
\mathcal M_{T2}=0.
\end{equation}
Alternatively, if the helicity axis is taken along one of the intermediate momenta $\mathbf q_i$ ($i=0,+,-$), then $\mathbf n\cdot\mathbf q_i=0$, so that
\begin{equation}
\mathcal M_L^{(i)}=0,\qquad
\mathcal M_{T1}^{(i)}=0,\qquad
\mathcal M_{T2}^{(i)}=i\,m_\phi\,|\mathbf n|\,\mathcal F,
\end{equation}
for $\hat{\mathbf e}_{T2}=\hat{\mathbf n}$. Therefore, independently of the basis choice, the amplitude is entirely controlled by the decay-plane normal vector.

\section{Angular distribution in $\phi\to\pi^+\pi^-\pi^0$}
In the $\phi$-meson rest frame, the decay
\begin{equation}
\phi(q)\to\pi^+(p_+)\pi^-(p_-)\pi^0(p_0)
\end{equation}
is described by
\begin{equation}
q^\mu=(m_\phi,\mathbf 0),\qquad p_i^\mu=(E_i,\mathbf p_i),\qquad \mathbf p_++\mathbf p_-+\mathbf p_0=0.
\end{equation}
The relative angle between the charged pions is
\begin{equation}
\cos\theta_{+-}=\frac{\mathbf p_+\cdot\mathbf p_-}{|\mathbf p_+|\,|\mathbf p_-|}=\frac{2m_{\pi^\pm}^2+2E_+E_- - s_{+-}}{2|\mathbf p_+|\,|\mathbf p_-|},
\end{equation}
with
\begin{equation}
E_i=\frac{m_\phi^2+m_i^2-s_{jk}}{2m_\phi},\qquad |\mathbf p_i|=\sqrt{E_i^2-m_i^2},
\end{equation}
and
\begin{equation}
s_{+-}+s_{+0}+s_{-0}=m_\phi^2+2m_{\pi^\pm}^2+m_{\pi^0}^2.
\end{equation}
Using the amplitude representation in the main text, the polarization-averaged squared amplitude is
\begin{equation}
\overline{|\mathcal M|^2}=\frac{m_\phi^2}{3}|\mathbf n|^2|\mathcal F(s_{ij})|^2.
\end{equation}
The differential decay width is then
\begin{equation}
d\Gamma=\frac{1}{3(2\pi)^3\,32m_\phi}|\mathbf n|^2|\mathcal F(s_{ij})|^2\,ds_{+-}\,ds_{+0}.
\end{equation}
Projecting onto $\cos\theta_{+-}$ gives
\begin{equation}
\frac{d\Gamma}{d\cos\theta_{+-}}=\frac{1}{3(2\pi)^3\,32m_\phi}\int ds_{+-}\,ds_{+0}\,|\mathbf n|^2|\mathcal F(s_{ij})|^2\,\delta\left[\cos\theta_{+-}-\cos\theta_{+-}(s_{+-},s_{+0})\right],
\end{equation}
where the integration is performed over the physical Dalitz region. This makes explicit that the angular distribution is a projection of the same underlying dynamics encoded in the Dalitz density. For the numerical results discussed in the main text, $\mathcal F$ in the above expressions is replaced by the FSI-improved amplitude $\mathcal F^{\rm FSI}$.


\end{document}